\documentclass[sigconf]{acmart}
\usepackage{algorithm}
\usepackage{algorithmic}
\AtBeginDocument{
 \providecommand\BibTeX{{%
 \normalfont B\kern-0.5em{\scshape i\kern-0.25em b}\kern-0.8em\TeX}}}

\usepackage{soul}

\renewcommand\footnotetextcopyrightpermission[1]{} % removes footnote with conference information in first column
\usepackage{pifont}% http://ctan.org/pkg/pifont

\usepackage{todonotes}

\usepackage{xcolor}

\usepackage{amsmath}

\newcommand\encircle[1]{%
\tikz[baseline=(X.base)] 
 \node (X) [draw, scale=0.75, shape=circle, inner sep=0, fill=black, text=white, minimum size=0em] {\strut #1};}

\begin{document}

\title{IMA-GNN: In-Memory Acceleration of Centralized and Decentralized Graph Neural Networks at the Edge}

\author{Mehrdad Morsali, Mahmoud Nazzal, Abdallah Khreishah and Shaahin Angizi}
\affiliation{
	\institution{Department of Electrical and Computer Engineering, New Jersey Institute of Technology, Newark, NJ 07102 \country{USA}\\
	}
}
\email{}

\begin{abstract}
In this paper, we propose IMA-GNN as an \underline{I}n-\underline{M}emory \underline{A}ccelerator for centralized and decentralized \underline{G}raph \underline{N}eural \underline{N}etwork inference, explore its potential in both settings, and provide a guideline for the community targeting flexible and efficient edge computation. Leveraging IMA-GNN, we first model the computation and communication latencies of edge devices. We then present practical case studies on GNN-based taxi demand and supply prediction and also adopt four large graph datasets to quantitatively compare and analyze centralized and decentralized settings. Our cross-layer simulation results demonstrate that on average, IMA-GNN in the centralized setting can obtain $\sim$790$\times$ communication speed-up compared to the decentralized GNN setting. However, the decentralized setting performs computation $\sim$1400$\times$ faster while reducing the power consumption per device. This further underlines the need for a hybrid semi-decentralized GNN approach. 
\end{abstract}

% <concept_id>10010583.10010600.10010607.10010608</concept_id>
% <concept_desc>Hardware~Dynamic memory</concept_desc>
% <concept_id>10010520.10010521.10010542.10010543</concept_id>
% <concept_desc>Computer systems organization~Reconfigurable computing</concept_desc>
% <concept_significance>300</concept_significance>

% \ccsdesc[500]{Hardware~Dynamic memory}
% \ccsdesc[300]{Computer systems organization~Reconfigurable computing}

\maketitle
\pagestyle{plain}

 \section{Introduction}
\begin{figure}[t]
\centering
\includegraphics [width=0.98\linewidth]{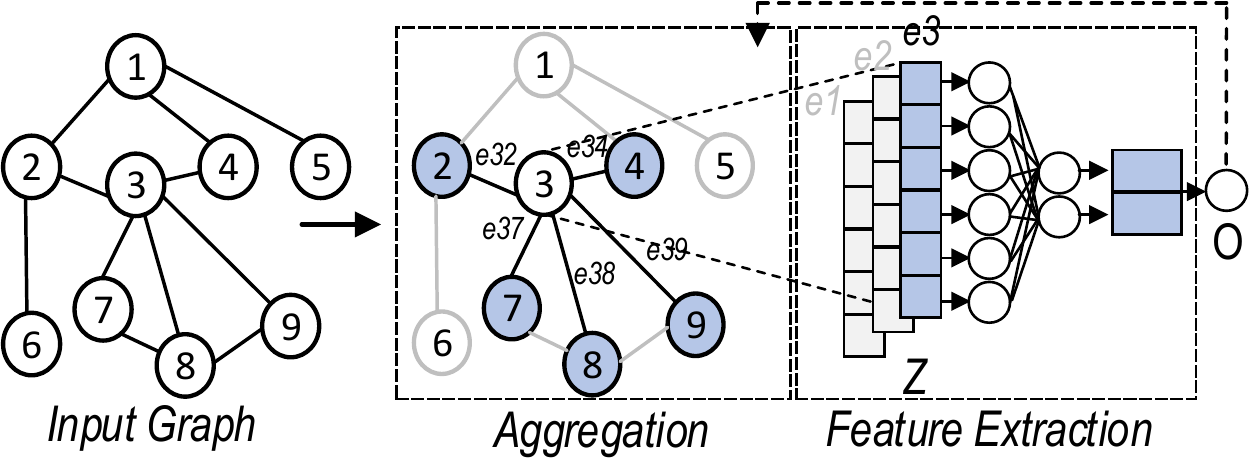} 
\vspace{-0.82em}
\caption{GNN's aggregation and feature extraction stages for a sample input graph.}
\vspace{-1.7em}
\label{GNN}
\end{figure}

Graph data structures appear naturally in many fields such as user accounts in a social network, atoms in a chemical molecule, and vehicles in a traffic system. Graph Neural Networks (GNN) extend deep learning to graph data by combining graph structure and node information through message passing and aggregation \cite{hamilton2017representation,sahu2017ubiquity}. This enables GNNs to deliver node embeddings to serve multiple downstream graph tasks such as node classification (inferring the class label of a node), link prediction (estimating the possibility of a link between given
nodes), and graph classification (inferring the class label of a graph). Essentially, GNNs obtain embeddings for the nodes in a given graph. Each node has a computational graph composed of its $k$-hop neighboring nodes. Node embeddings are obtained by alternating between message passing, i.e., communicating local information across nodes, and aggregation where received messages along with previous node information are used to obtain an updated embedding. Message passing is done according to the graph structure, whereas aggregation is done by the (trainable) neural network layers of the GNN model \cite{kiningham2022grip,yan2020hygcn}. 
As shown in Fig. \ref{GNN}, for a sample input graph G, first, in the {aggregation} stage, each node aggregates the information from all neighbors (nodes 2, 4, 7, 8, 9) with its own data (node 3) and creates the $\boldsymbol{Z}$ matrix that represents the aggregated node features from node 3 and its neighbors. During the {feature extraction} stage, the result of the aggregation stage ($\boldsymbol{Z}$) is fed into a Multi-Layer Perceptron (MLP) or a Convolutional Neural Network (CNN) model to generate the output matrix indicated by $\boldsymbol{O}$. The GNN workflow performs the same steps for all the nodes in the graph.

% Consequently, GNNs have achieved state-of-the-art performances in a wide variety of applications, such as social network analysis \cite{li2017deepcas}, fraud detection \cite{dou2020enhancing}, and traffic forecasting \cite{jiang2022graph}. 

Graphs in many real-world application areas are naturally enormous. For example, social media and e-commerce graphs such as Facebook and Amazon graphs have billions of nodes and edges. Furthermore, local node features are typically of large dimensions ranging between hundreds to several thousands \cite{hu2020open}. This creates a corresponding time and memory space demand for training and testing on the underlying GNN models \cite{zheng2020distdgl,zhou2020graph}. This is because nodes are mutually dependent, and therefore, graphs can not be arbitrarily divided into smaller subgraphs. Techniques such as {neighborhood sampling} \cite{hamilton2017inductive} may help to some extent, but depending on the graph structure, even a sampled computation graph and associated features may not fit in the memory of a single GPU. Besides, off-chip memory access is a critical issue in the Von-Neumann computing architecture \cite{mandal2022coin,li2021gcnax,abedin2022mr,angizi2022pisa} and to minimize the latency and power consumption, the Processing-in-Memory (PIM) accelerators have been set forth focusing on centralized GNNs. 
Along this line, HyGNN \cite{yan2020hygcn} supports hybrid computing and memory access patterns and performs GCN computations efficiently using a hybrid architecture. Utilizing two dedicated processing engines, HyGNN tackles irregularity with an aggregation engine and leverages regularity with a combination engine. The irregularity arises from the topology-dependent aggregation process which is inherently random and sparse, while the transformation process is a neural network operation that has a static and regular execution pattern.
In the same way, GRIP \cite{kiningham2022grip} divides the GNN computations into different engines. Multiple parallel prefetch and reduction engines have been used for aggregation to alleviate the irregularity. For regular computation and memory access patterns, GRIP utilizes a high-performance matrix multiply engine with dedicated memory. Another edge-centric paradigm, EnGN \cite{liang2020engn}, has been also implemented using a ring-edge-reduce dataflow. EnGN alleviates the poor locality of sparse and random connected vertices (nodes). AWB-GCN \cite{geng2020awb} proposes a hardware-based workload distribution auto-tuning framework consisting of three workload rebalancing techniques to alleviate the extreme workload imbalance. PIM-GCN \cite{challapalle2021crossbar} presents a node-stationary dataflow with support for compressed sparse row and column graph representations.
% \Mahmoud{May be we can mention some limitations of these approaches} Excellent Comment, we must consider this, however in this work due to lack of time and space, we didn't compare our work with other centralized designs. just centralized vs decentralized

\begin{figure*}[t]
\centering
\includegraphics [width=0.99\linewidth]{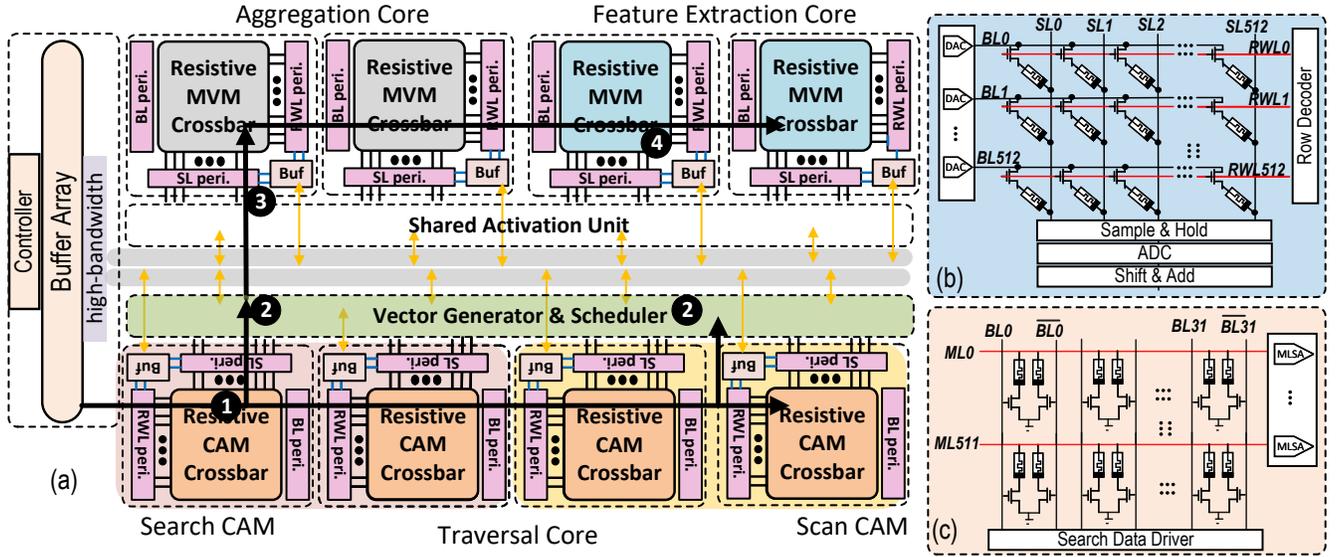} 
\vspace{-0.75em}
\caption{(a) The proposed IMA-GNN architecture with resistive CAM traversal core, resistive MVM aggregation, and feature extraction cores, (b) Resistive MVM crossbar, (c) Resistive CAM crossbar.}
\vspace{-0.2em}
\label{arc}
\end{figure*}

Distributed (decentralized) GNN training and inference \cite{li2020graph,tolstaya2020learning} is based on dividing a given graph into smaller subgraphs that can be more easily processed with distributed devices. Despite its ability to mitigate the computation overhead by load sharing, decentralized GNN operation faces a major bottleneck; the excessive communication overhead between nodes in different distributed devices \cite{zheng2020distdgl}.
To the best of our knowledge, this work is among the first to explore and compare in-memory acceleration in both centralized and decentralized GNN settings and to offer a design guideline to the community. 
The main contributions of this paper are as follows:
(1) We develop a PIM architecture with RRAM arrays based on a set of innovative micro-architectural designs that can be optimized and used for centralized and decentralized GNN inference for efficiency and speed-up; (2) We model the latency and power consumption of GNN accelerators implemented in centralized and decentralized settings considering the computation and communication between edge devices; and (3) We present a solid bottom-up evaluation framework to analyze the performance of the whole system in real scenarios and through adopting large graph datasets.
\vspace{-0.5em}

\section{Proposed IMA-GNN}
\subsection{Architecture Overview}
The IMA-GNN is a high-performance and energy-efficient RRAM crossbar-based accelerator developed to execute GNN's pivotal operations in both centralized and decentralized settings inspired by \cite{challapalle2021crossbar}. As shown in Fig. \ref{arc}(a), IMA-GNN comprises three computation cores, i.e., traversal, aggregation, and feature extraction as well as peripherals such as a buffer array and a controller. The traversal core consists of resistive Content Addressable Memory (CAM) crossbars capable of search and comparison operations (Fig. \ref{arc}(c)). All resistive CAM crossbars on the bottom side are connected to a shared vector generator \& scheduler unit and then to a high-bandwidth bus to communicate with other cores at the top. The aggregation core includes resistive crossbars (Fig. \ref{arc}(b)) to perform in-situ Matrix-Vector-Multiplication (MVM) operations for the feature aggregation. The feature extraction core is designed with a similar resistive crossbar but a different size to take care of transformation in GNN inference. The resistive MVM crossbars are connected to a shared activation unit.\vspace{-0.5em}

\subsection{MVM \& CAM Crossbars}
The RRAM crossbar memory arrays are widely explored as a potential parallel engine to execute MVM operation and scan and search \cite{angizi2022pisa,challapalle2021crossbar,abedin2022mr}. As shown in Fig. \ref{arc}(b) in the MVM crossbars, the weight parameters are first stored as resistance states in each RRAM device in a 1-Transistor-1-RRAM (1T1R) structure, and then the input binary bit-strings, as the inputs to the crossbar array, are converted by the Digital-to-Analog Converter (DAC) into voltages $V_i$ and applied to Bit-Lines ($BL$s) in parallel. The weighted currents generated from the RRAM cells sharing a Source-Line ($SL$) are accumulated resulting in an intrinsic dot-products operation. The accumulated values are then sampled by Sample \& Hold unit and then converted to binary data using Analog-to-Digital Converters (ADC). The partial-product results from each $SL$ are further processed by the Shift \& Add unit to generate the final result. As shown in Fig. \ref{arc}(c), in the CAM crossbars, each Ternary CAM (TCAM) cell consists of 2-Transistor-2-RRAM (2T2R) to accomplish the XNOR search operation on each pair of cells. For this operation, $BL$ and $\overline{BL}$s are valued with the search data by the Search Data Driver. Accordingly, the Sense Amplifier connected to Match-Lines ($ML$SA) senses whether the row is a match or mismatch with the reference connected to Vdd. In the compare operation, $BL$s are grounded and $\overline{BL}$s are connected to increasing calibrated voltages from the Least Significant Bit (LSB) to the Most Significant Bits (MSB).

\subsection{Accelerator Dataflow}
\begin{figure}[h]
\centering
\includegraphics [width=0.99\linewidth]{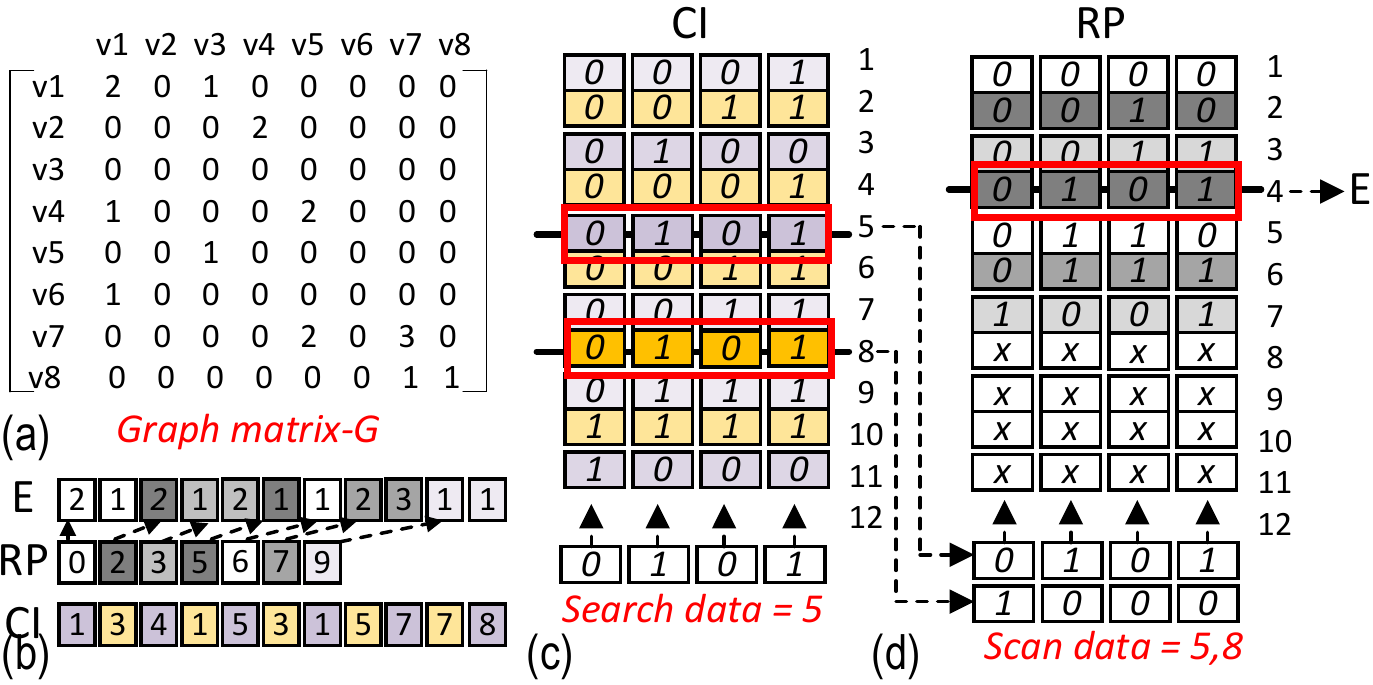} 
\vspace{-0.75em}
\caption{IMA-GNN's hardware mapping and acceleration in traversal core: (a) Sample graph adjacency matrix, (b) CSR representation, (c) Search CAM operation, (d) Scan CAM operation.}
\vspace{-1em}
\label{mapping}
\end{figure}

Once the edge buffers shown in Fig. \ref{arc}(a) on the left have been loaded with graph data in either centralized or decentralized GNN settings, the traversal core starts processing edges. The traversal core performs two essential CAM-based operations, i.e., search and compare. To maximize the data reuse of feature data in IMA-GNN, the traversal core implements an efficient node-stationary dataflow by buffering a set of node features in the buffer array and reusing it for the aggregation core.
IMA-GNN leverages a Compressed Sparse Row (CSR) format \cite{lu2019redesk} to form the Edge weight array (E), Column Index array (CI), and Row Pointer array (RP) and loads the graph data to search and scan CAMs (Fig. \ref{arc}(a) \encircle{1}). A sample graph adjacency matrix and the corresponding CSR format are shown in Fig. \ref{mapping}(a)-(b).
Any destination node then operates as an input to the search CAM as shown in the data mapping in Fig. \ref{mapping}(c) and rows that match the search data are activated. Matching rows are reference inputs for comparison in the scan CAM, which determines the source nodes with edges to the destination node by comparing the input row with RP (Fig. \ref{mapping}(d)). Next, the vector generator \& scheduler unit receives the result of scan CAM and edge data to render input control vectors for the aggregation core (Fig. \ref{arc}(a) \encircle{2}). This will activate particular rows of resistive aggregation core corresponding to incoming edges.
Next, the aggregation core input buffers receive input vectors for each destination node along with the destination node. The aggregation core starts with source node features or feature dimensions across its own cluster \encircle{3}. IMA-GNN is equipped with double buffering for feature data and graph data. This feature enables overlapping writing/programming phases and the traversal stage. 
Next, the updated destination node features are fed to the feature extraction core's crossbars programmed with weights \encircle{4}. Besides, similar to \cite{mandal2022coin}, to maximize the crossbar utilization in the aggregation core, both the aggregation and feature extraction cores could work in parallel. 
% \Mahmoud{They do work in parallel, right?} YES, absolutely

\section{Network Modeling}
\begin{figure}[t]
\centering
\includegraphics [width=0.99\linewidth]{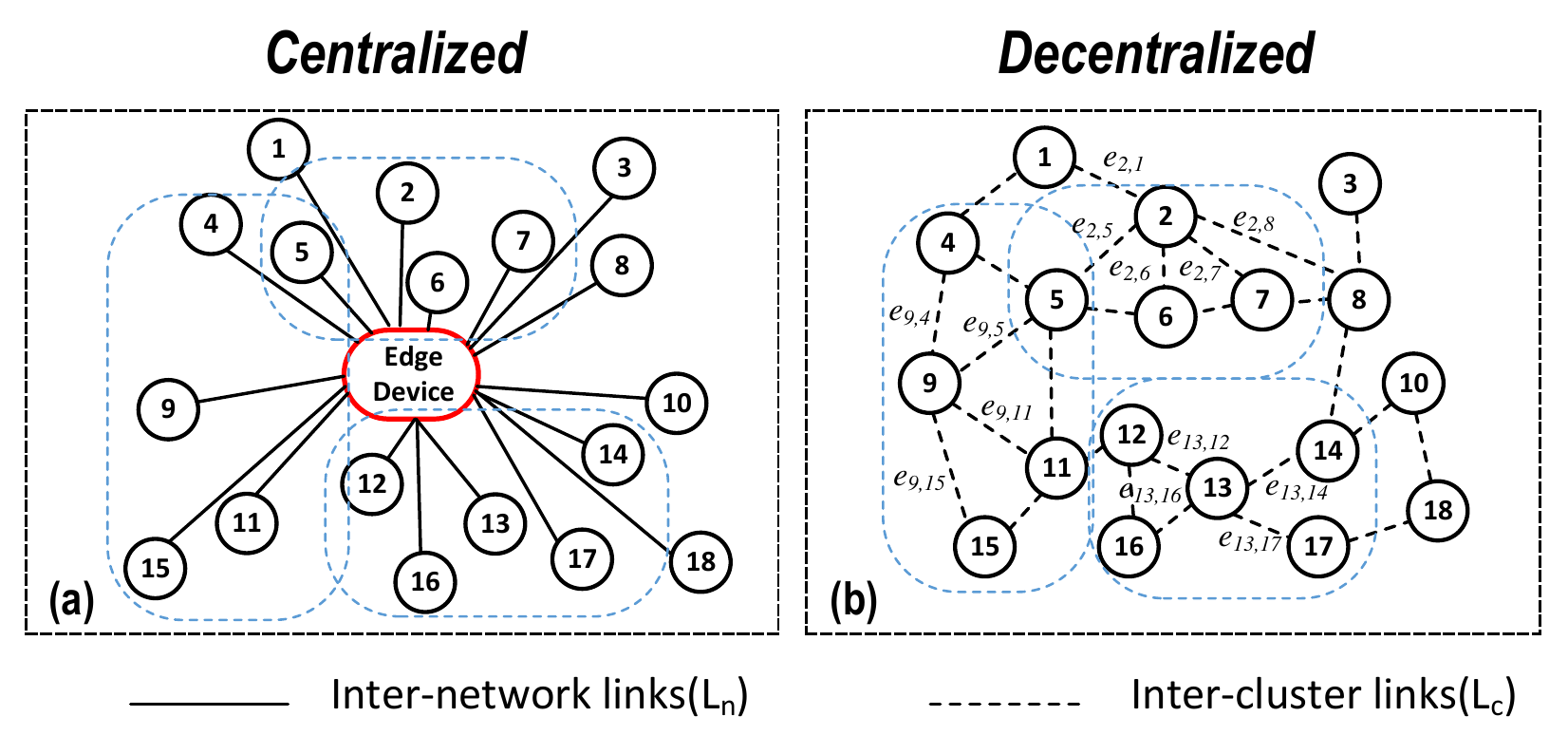} 
\vspace{-0.75em}
\caption{Intra- and inter-edge links in a sample (a) centralized versus (b) decentralized GNN.}
\vspace{-1em}
\label{edge}
\end{figure}

We explore both centralized and decentralized GNN landscapes to fairly model the performance metrics in both settings. Figure \ref{edge} shows a sample graph with $N$ ($N\in\mathbb{Z}^+$) nodes (edge devices). In the centralized GNN setting, a single powerful node as the accelerator is designed with embedded traversal, aggregation, and feature extraction cores to communicate through fast inter-network links ($L_n$) \cite{mannoni2019comparison} to aggregate all edge devices' information and handle the computation burdens of transformation. \textcolor{black}{These cores have $M_1$, $M_2$, and $M_3$ times larger allocated computing hardware for traversal, aggregation, and feature extraction operations respectively than a single node in the decentralized mode. Thus, we assume the processing capability of the edge device in the centralized setting is $M_1$, $M_2$, and $M_3$ times larger than the processing capability of a single node in the decentralized mode in the traversal, aggregation, and feature extraction operations, respectively.}
In the decentralized GNN setting, each edge device is observed as an accelerator with reduced traversal and aggregation cores and in addition to a copy of our network, has an embedded feature extraction core processing $L$ layers. The output of the feature extraction core at each edge device is only communicated to the adjacent edge devices at a defined cluster as shown in Fig. \ref{edge}(b). Therefore, the communication between neighbors through inter-cluster links ($L_c$) \cite{miya2021experimental} generates a communication volume as well. The bidirectional communication volume between node-$i$ to node-$j$ is represented as $e_{i,j}$. Therefore, the minimization of the accelerator's computational latency/power and communication latency/power is a pivotal need in the research community.
We estimate the centralized and decentralized GNN accelerators' latency as:

\begin{equation}
 \small T_{Net}(N)=T_{compute}(N)+T_{communicate}(N).
\end{equation}
In an \textit{N}-edge device graph shown in Fig. \ref{edge}, denoting by $t_1$, $t_2$, and $t_3$ the traversal, aggregation, and feature extraction cores' latency, respectively, the computation latency of a single node in the decentralized GNNs can be estimated by:
\begin{equation}
 \small T_{compute-decentralized}=t_1+t_2+t_3,
\end{equation}

\noindent where in the centralized setting, considering the processing capability of a single powerful edge device the computation latency is given by:
\begin{equation}
 \small T_{compute-centralized}=({{t_1}/M_1}+ {{t_2}/M_2} + {{t_3}/M_3})\times (N-1).
\end{equation}
In the decentralized setting, the communication latency, $T_{communicate}$, can be given by:
\begin{equation}
 \small T_{communicate-centralized}=(t_e+(c_s\times t(L_c)))\times2 ,
\end{equation}
where $t_e$ is the required time for establishing a connection between two adjacent nodes, $c_s$ denotes the number of adjacent nodes inside a cluster, $t(L_c)$ denotes the latency of the inter-cluster link, and number 2 is to model a two-way link.
We assume that data communication inside each cluster is done in a sequential way, thus the number of adjacent nodes is multiplied by the latency of the inter-cluster link.
For the centralized setting, we assume data transfer between the central edge device and nodes is done in a concurrent way. Therefore, the communication latency, $T_{communicate}$, for centralized inference can be given by: 
\begin{equation}
 \small T_{communicate-decentralized}=t(L_n), 
\end{equation}
where $t(L_n)$ is the latency of the inter-network link. 
Suppose each edge device runs a GNN with \textit{X}-layers, the number of input and output neuron activations for a layer \textit{$x$} for 1$\leqslant x\leqslant$X can be given by $\alpha(x)$ and $\alpha(x+1)$, respectively. 
The total power consumption of GNNs implemented in the proposed accelerator can be developed as: 
\begin{equation}
 \small P_{Net}(N)=P_{compute}(N)+P_{communicate}(N).
\end{equation}

The first part accounts for the computation power and the second part considers communication power for inter-network and inter-cluster links. In the centralized setting, $P_{compute-centralized}$ is given by $\frac{E_{compute-centralized}}{T_{compute-centralized}}$ and $P_{communicate-centralized}$ can be given by $p(L_n)\times2$. Here $p(L_n)$ denotes the power consumption of the inter-network link and number 2 is to model a two-way transfer. $E_{compute-centralized}$ is readily calculated by achieving the energy consumption values of traversal, aggregation, and feature extraction cores. As for the decentralized setting, $P_{compute-decentralized}$ can be computed with respect to energy and latency parameters. 
We consider $\sum_{n=1}^{c_s}p_{n}(c_s(n)(c_s(n)-1))$ transactions between all accelerators inside the cluster. Considering the X-layer GNN, $P_{communicate}$ can be expressed as follow:
\begin{equation}
 \small P_{communicate-decentralized}= \frac{1}{t(L_c)} \times\sum_{x=1}^{X-1}\alpha(x+1)\times E_{perBit}
\end{equation}

% Note that, in an ideal decentralized networks always \textit{m=N} that means \textit{T}=0.
% The third part of the objective function accounts for the inter-edge communication power. 
% Assume $p_{w,z}$ represents the probability of having a connection between two nodes-\textit{w} and \textit{z} located in different edges, there is a total $T_{w,z}=p_{w,z}(\frac{N}{m})(\frac{N}{m})$ transactions between two edges. Since each node generates $\alpha(x+1)$ output activation bits after processing the $X^{th}$ between two edges, the total inter-edge communication is achieved by the summations of all edge pairs.
% \begin{equation}
% \small E_{interEdge}(m)=\sum_{w=1}^{m}\sum_{z=1}^{m}T_{w,z}\times\sum_{x=1}^{X-1}\alpha(x+1)\times E_{perBit}
% \end{equation}

\section{Experiments}

\subsection{Evaluation Framework}
\begin{figure}[h]
\begin{center}
\begin{tabular}{c}
\includegraphics [width=0.99\linewidth]{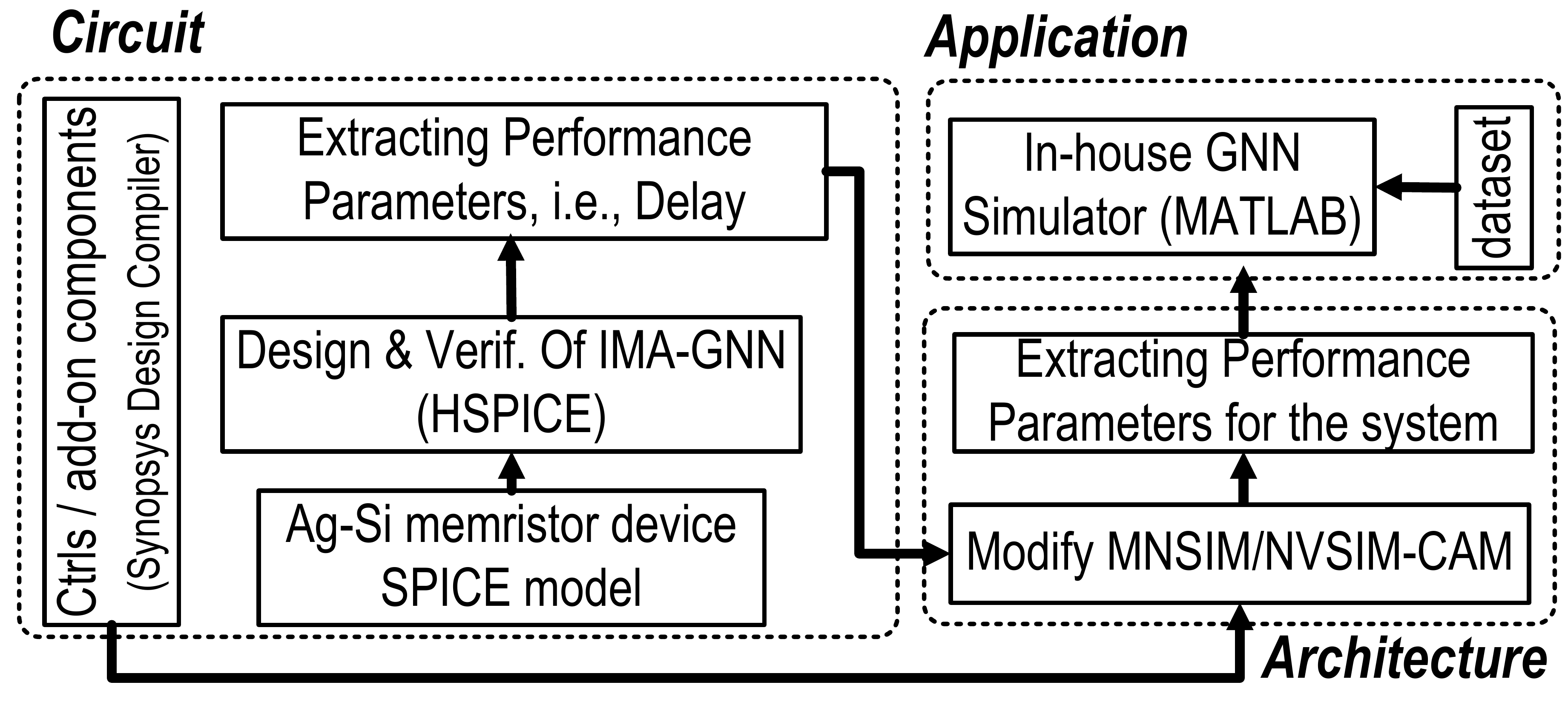}\vspace{-0.4em}
 \end{tabular} \vspace{-0.7em}
\caption{Circuit-to-application evaluation framework.}\vspace{-1em}
\label{framework}
\end{center}
\end{figure}

To evaluate the performance of the proposed architecture, a comprehensive bottom-up evaluation framework is developed as depicted in Fig. \ref{framework}. At the circuit-level, we use the SPICE model for memristors with the Ag-Si memristor device parameters from \cite{gao2012analog}. We then combine the SPICE models of CMOS transistors and memristors under NCSU 45nm CMOS PDK \cite{NCSU_PDK} to fully design and verify the IMA-GNN cores in HSPICE and to extract performance parameters such as delay and power consumption. We use the Synopsys Design Compiler \cite{DC} to develop the controller and buffer array using a standard industry-level 45nm technology. At the architecture-level, we modify and configured the NVSIM-CAM \cite{li2016nvsim} memory evaluation tool and MNSIM \cite{zhu2020mnsim} with our circuit-level results to extract the performance parameters for traversal, aggregation, and feature extraction cores. The results are then fed to an in-house MATLAB code with the graphs taken as input to calculate the estimated latency and power consumption for various workloads.
In this study, to have a fair comparison between GNN settings, we set up IMA-GNN's traversal, aggregation, and feature extraction cores with 2K$\times$(512$\times$32), 1K$\times$(512$\times$512), and 256$\times$(128$\times$128), respectively, for the centralized setting and 512$\times$32, 512$\times$512, and 128$\times$128, respectively, for the decentralized setting. It is noteworthy that in addition to the size and number of crossbars in each core, there are several factors that determine the total latency and power consumption of IMA-GNN in each setting, such as the distribution of graph edges across nodes, the availability of graph data, on-chip storage, and off-chip data accesses.

\subsection{Traffic Demand Forecasting}
\begin{figure}[b]
\centering
\resizebox{0.99\columnwidth}{!}{
\begin{tabular}{cc}
\includegraphics[width=10cm]{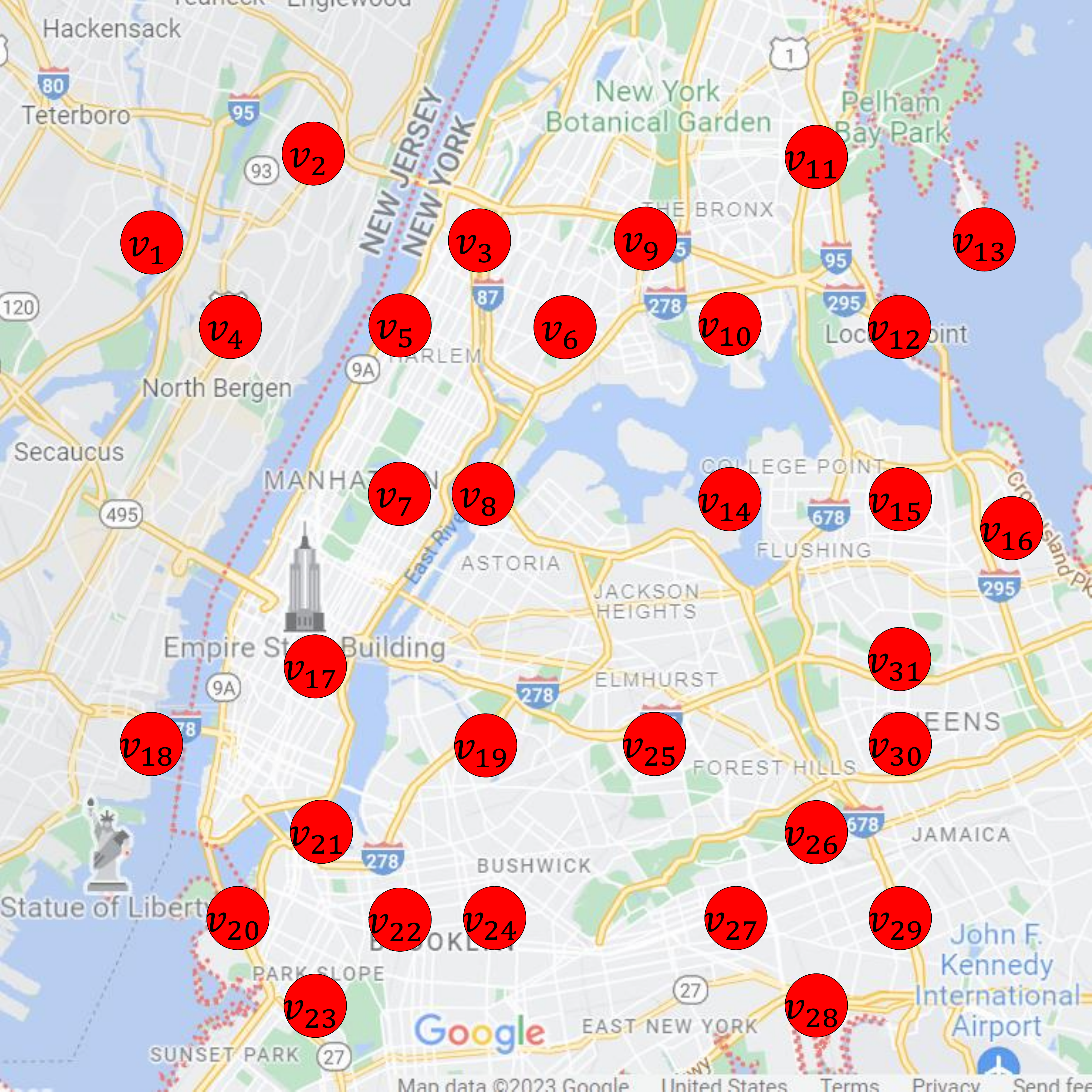}
&
\includegraphics[width=10cm]{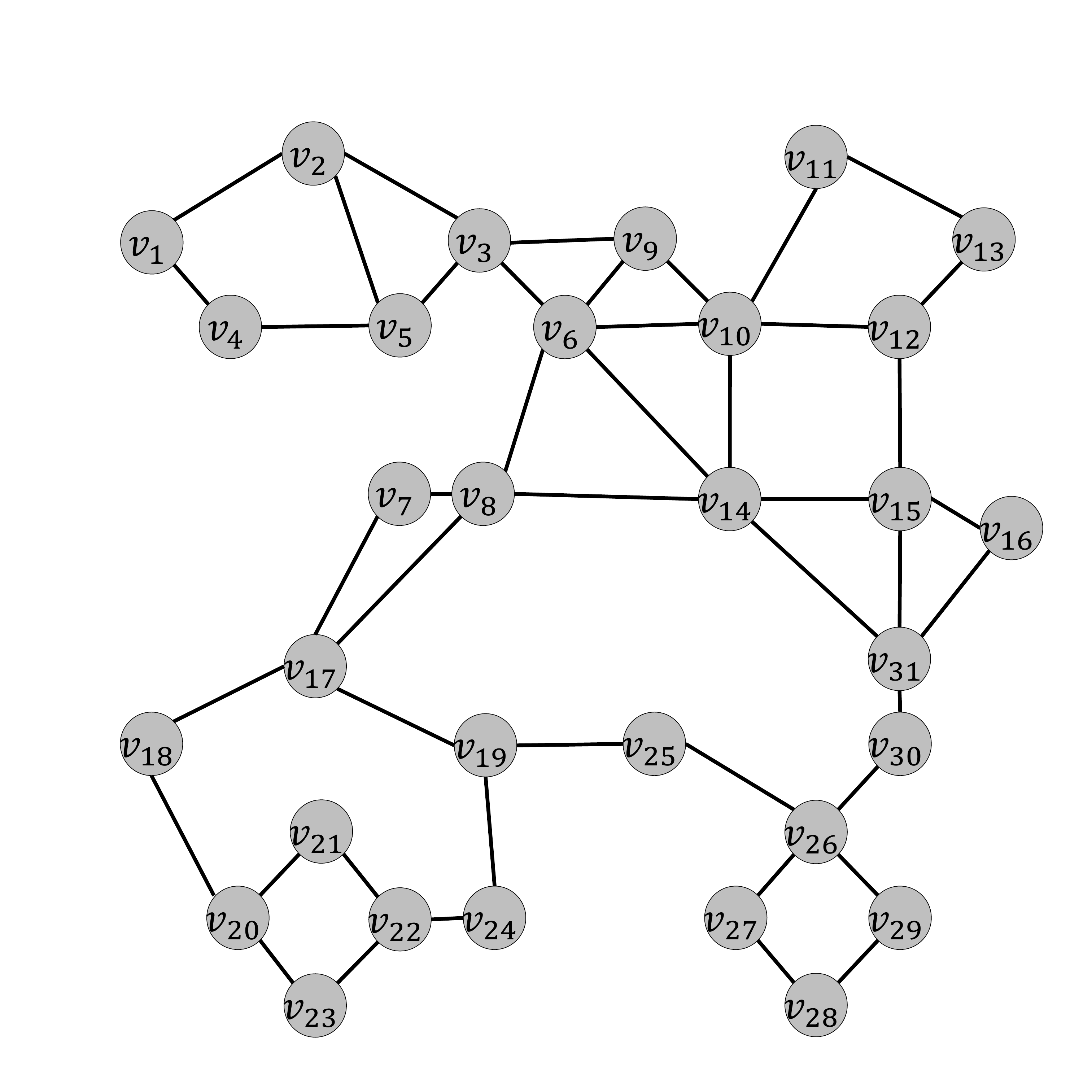}
\\
\Huge{(a)}
&
\Huge{(b)}
\end{tabular}}
\caption{(a) Taxi representation as a graph, and (b) Decentralized GNN operation.}
\label{graph_construction} 
\end{figure}

\begin{figure}[t]
\centering
\resizebox{0.99\columnwidth}{!}{
\includegraphics[width=12cm]{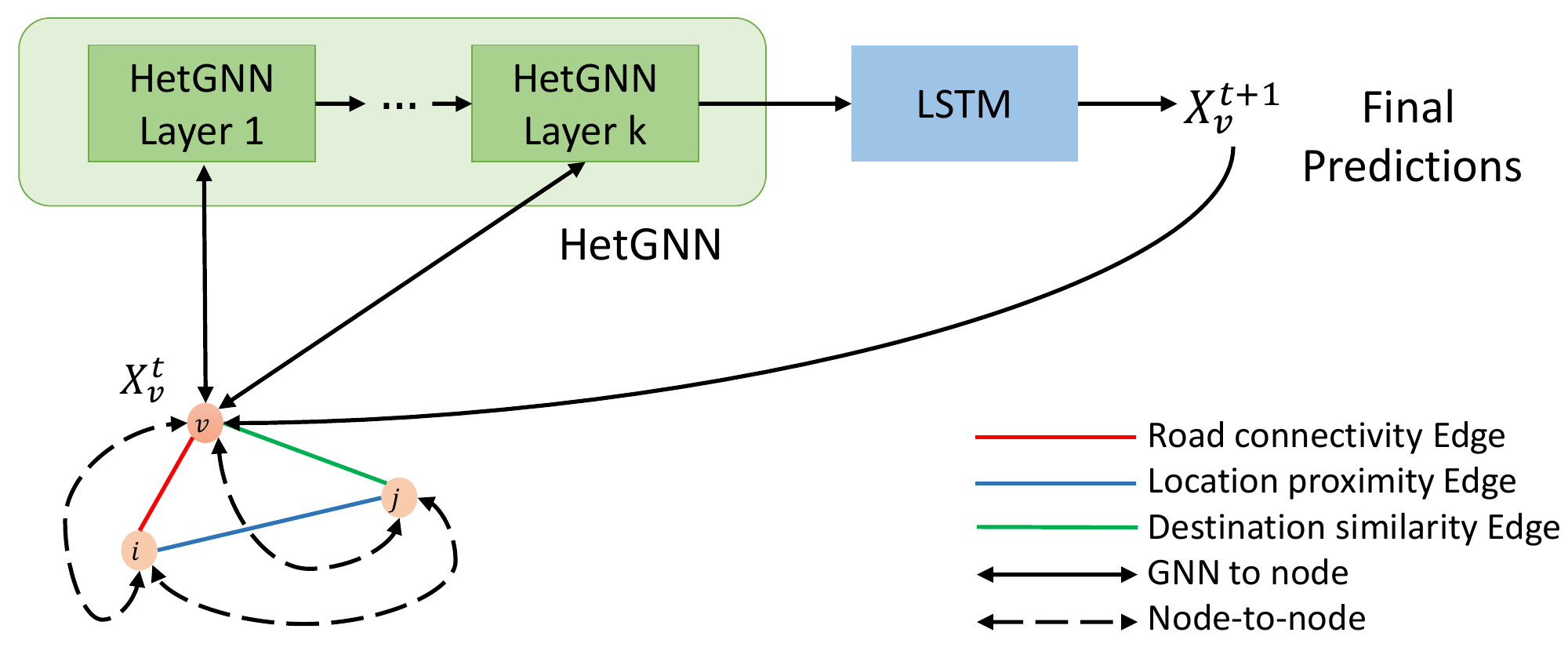}}
\caption{The architecture of the hetGNN-LSTM based prediction in \cite{nazzal2023traffic}.}
\label{system_architecture} 
\end{figure}

As a case study on the potential of hardware-accelerated decentralized GNNs in real-world applications, we pick a recent work on city-wide multi-relational and spatiotemporal taxi demand and supply forecasting
% A preliminary version of that work is published in the pre-print 
\cite{nazzal2023traffic}. 
Figure \ref{graph_construction}(a) and (b) show sample taxis in a city region and their corresponding graph representation, respectively. This graph is composed of taxi nodes linked by three edge types; \textit{road connectivity}, \textit{location proximity}, and \textit{destination similarity} edges linking taxis connected by a road, being nearby, and targeting nearby destinations, respectively. For each taxi node, the objective is to predict the values of transportation demand and supply for a region surrounding it. This is done based on both historical demands and supplies in the node's surrounding region and the corresponding messages shared by other connected nodes (taxis). Formally, the objective is to train a GNN operator $\mathcal{F}$ that predicts the $Q$-long future passenger demands and supplies in an $m$ by $n$ region surrounding each taxi at a time instant $t$, $\mathbf{X}_{t+1: t+Q} \in \mathbf{R}^{Q \times m \times n}$. This is based on the $P$ historical values of the demands and supplies in this region, and the corresponding $P$ historical values passed from the $k$-hop neighboring nodes. So, $
\left[\mathbf{X}_{t-P+1: t}, G\right] \stackrel{\mathcal{F}}{\longrightarrow} \mathbf{X}_{t+1: t+Q},
$
where $\mathbf{X}_{t-P+1: t} \in \mathbf{R}^{P \times m \times n}$, and $G$ denotes the node's computational graph. Due to the existence of multiple edge types and the time-dependency, $\mathcal{F}$ is composed of a heterogeneous GNN (hetGNN) used for message passing, followed by a Long-Short-Term Memory (LSTM) network to incorporate time dependency, as depicted in Fig. \ref{system_architecture}.

\par The enormous sizes of transportation graphs 
make it challenging to apply the model in \cite{nazzal2023traffic} in centralized GNN inference. This limitation is further aggravated by the hetGNNs and the huge volumes of local node information. As a remedy to resolve this limitation, the authors in \cite{nazzal2023traffic} propose a decentralized GNN inference approach. In this approach, each taxi node has a copy of the model (hetGNN-LSTM), exchanges messages with its $k$-hop neighbors, and then uses the hetGNN-LSTM model to predict the demands and supplies in its surrounding region. A natural advantage of this approach is handling dynamically varying graph structures. Nevertheless, despite the promising advantages of decentralization, there is still a demanding need for reducing the overall computation and communication latency in the operation of the model.

%%Communication latency numbers from Mahmoud's Document
\par In this experiment, for the centralized GNN setting, the overall latency (in terms of transmission delay) for sending and receiving a packet of 300 Bytes is considered 1.1 ms where the range of the network is 300 meters \cite{mannoni2019comparison}. This latency is the average overall latency to correctly receive a packet of 300 bytes. Thus, for a packet size of 864 bytes, which is the size of our data, the overall transmission delay can be $\sim$3.3 ms.

As for the decentralized GNN setting, we assume that nodes in the graphs of \cite{nazzal2023traffic} communicate with each other using an ad-hoc wireless network that uses channel 9 (2.452 GHz) of IEEE 802.11n, where the transmission power is fixed to –31 dBm, and bandwidth is 20 MHz. In this configuration, source nodes feed their messages to nearby proxy (relay) nodes which forward the messages to the next nodes, and so on. Since source nodes have more computation compared to proxy (relay) nodes, they incur more delay.
We leverage the proposed bottom-up evaluation framework to estimate IMA-GNN architecture performance in both centralized and decentralized settings. In our evaluation, the number of nodes (taxis) of the graph and the cluster size ($c_s$) are set as 10000 and 10, respectively. The evaluation results of the latency and power consumption are tabulated in Table \ref{taxilatency}.
\begin{table}[h]
\centering
\caption{Computation and communication latency/power of IMA-GNN accelerator.} \vspace{-1em} 
\scalebox{0.85}{
\begin{tabular}{|l|cc|cc|l}
\cline{1-5}
Settings & \multicolumn{2}{c|}{Centralized} & \multicolumn{2}{c|}{Decentralized} & \\ \cline{1-5}
Figure of merits & \multicolumn{1}{c|}{Latency} & Power & \multicolumn{1}{c|}{Latency} & Power & \\ \cline{1-5}
Traversal & \multicolumn{1}{c|}{38.43 ns} & 10.8 mW & \multicolumn{1}{c|}{7.68 ns} & 0.21 mW & \\ \cline{1-5}
Aggregation & \multicolumn{1}{c|}{142.77 $\mu$s} & 780.1 mW & \multicolumn{1}{c|}{14.27 $\mu$s} & 41.6 mW & \multicolumn{1}{c}{} \\ \cline{1-5}
Feature extraction & \multicolumn{1}{c|}{14.53 $\mu$s} & 32.21 mW & \multicolumn{1}{c|}{0.37 $\mu$s} & 3.68 mW & \multicolumn{1}{c}{} \\ \cline{1-5}
Computation (Net) & \multicolumn{1}{c|}{157.34 $\mu$s} & 823.11 mW & \multicolumn{1}{c|}{14.6 $\mu$s} & 45.49 mW & \\ \cline{1-5}
Communication & \multicolumn{1}{c|}{3.30 ms} & - & \multicolumn{1}{c|}{406 ms} & - & \\ \cline{1-5}
\end{tabular}
}
\label{taxilatency}
\end{table}
 
\par In view of the results, the decentralized setting in \cite{nazzal2023traffic} improves the total computation latency by a factor of $\sim$10$\times$. This is achieved by reducing the traversal latency, aggregation latency, and feature extraction latency by factors of 5$\times$, 10$\times$, and $\sim$39$\times$, respectively. Thus, a huge improvement can be observed in terms of computation latency. However, in terms of communication, the centralized setting acts much better than the decentralized setting by incurring a $\sim$120$\times$ less latency.
As for computation power consumption, we observe that the decentralized setting reduces the power budget per node by a factor of 18$\times$. The aggregation core of IMA-GNN consumes most of the power in both centralized and decentralized settings as well as the highest latency.
Overall, each of the settings has its own advantage from a different point of view. In the next subsection, more graph datasets with different characteristics are studied in order to further elucidate the pros and cons of centralized and decentralized settings. 

\subsection{Graph Datasets}
In the second case study, four graph datasets, LiveJournal, Collab, Cora, and Citeseer \cite{yan2019alleviating,challapalle2021crossbar} are used to evaluate the inference latency using the proposed IMA-GNN in centralized and decentralized settings. Key graph statistics of these datasets are provided in Table \ref{datasets}. A given vertex is mapped deterministically to a fixed-sized, uniform sample of its neighbors.

\begin{figure}[t]
\centering
\resizebox{0.99\columnwidth}{!}{
\includegraphics[width=12cm]{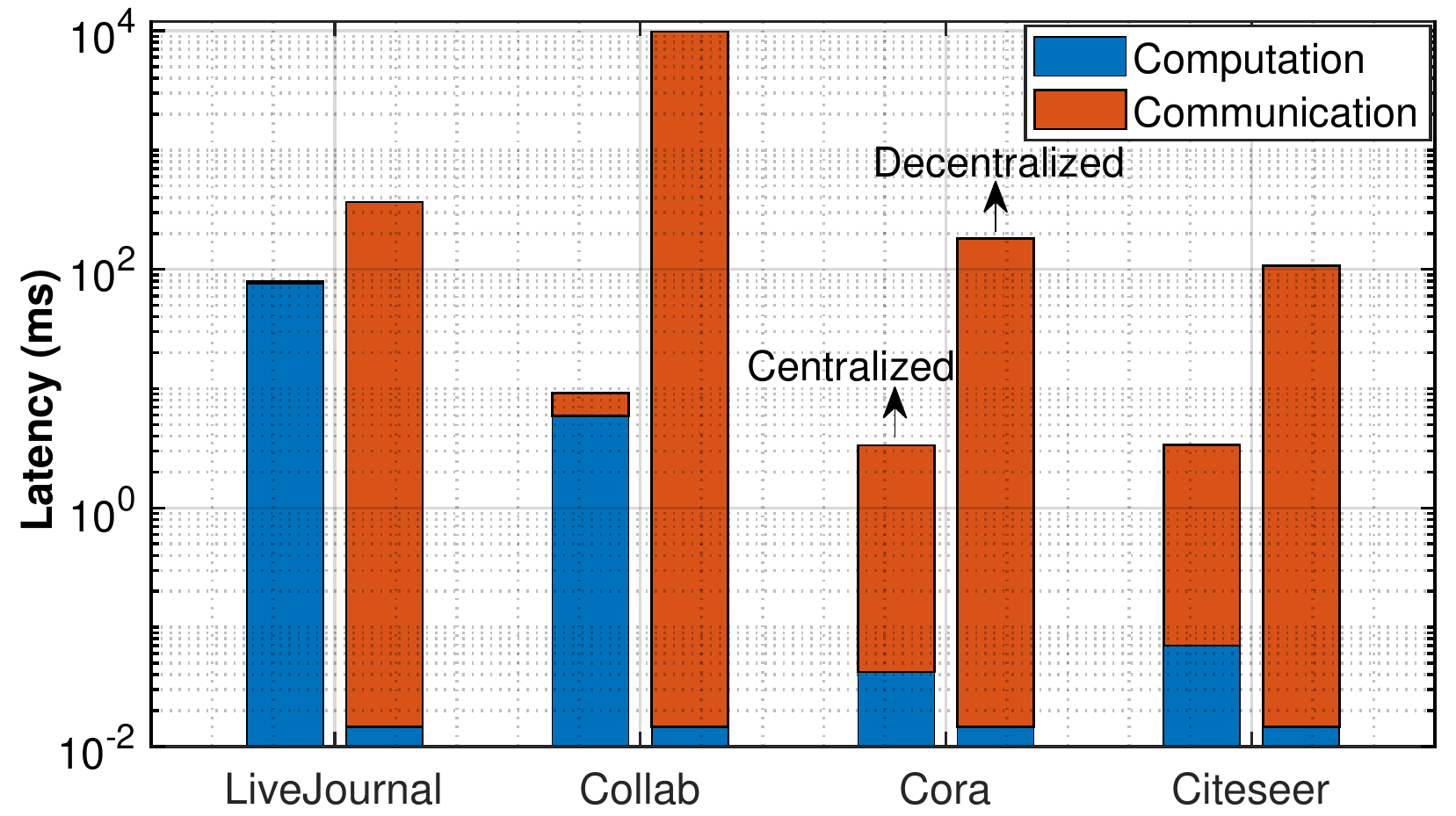}}
\caption{Breakdown of communication and computation latency of under-test graph datasets for centralized and decentralized settings.}\vspace{-1em}
\label{bargraph}
\end{figure}

\begin{table}[b]
\caption{Key statistics of the graph datasets used.} \vspace{-1em} 
\scalebox{0.85}{
\begin{tabular}{|l|c|c|c|c|}
\hline
{Datasets} & LiveJournal & Collab & Cora & Citeseer\\ \hline
{Number of Nodes} & 4,847,571 & 372,475 & 2708 & 3,327\\ \hline
{Number of Edges} &  68,993,773 & 24,574,995 &  5429 & 4,732\\ \hline
{Feature Length} & 1 & 496 & 1433 & 3,703\\ \hline
{Average $C_s$ }  & 9 & 263 & 4 & 2\\ \hline
\end{tabular}}
\label{datasets}
\end{table}
Figure \ref{bargraph} shows the latency for the four aforementioned graph datasets. Each bar consists of two parts; computation latency and communication latency. For each dataset, we have two bars representing the latency in the centralized (left) and decentralized (right) settings. 
%In view of Fig. \ref{bargraph}, amongst the four datasets, LiveJournal has the largest latency both in the centralized and decentralized settings due to its large graph size.
By close observation, it can be realized that in all under-test datasets, the computation latency of the decentralized setting is less than that of the centralized setting. Especially, the difference between the computation latency of centralized and decentralized settings is huge in LiveJournal and Collab datasets, where the graph size is much larger than the other two datasets. In view of Fig. \ref{bargraph}, amongst the four datasets, LiveJournal has the largest computation latency in the centralized settings because it owns the largest number of nodes. This is because in the decentralized mode, as each node is responsible to do its' own computation, the computation latency is independent of the total number of nodes and doesn't increase when the number of nodes grows. Conversely, in the centralized setting, by growing the number of nodes, the computation burden of the edge device will increase, causing an increment in the computation latency. On average for these four datasets, the decentralized setting performs computations $\sim$1400$\times$ faster. However, the communication latency which is the dominant part of the total latency is much higher in the decentralized setting as each node is required to establish a peer-to-peer connection and transfer data with all adjacent nodes sequentially. According to Fig. \ref{bargraph}, Collab has the largest communication latency amongst the four datasets, in the decentralized settings due to its large Average $C_s$, where each node is required to communicate with a large number of adjacent nodes sequentially. However, in the centralized mode, all nodes are connected to the edge device using a fast and mature connection and can transfer data in a parallel way. In terms of the communication latency, for four under-test datasets, the centralized setting is $\sim$790$\times$ faster than the decentralized setting.

In addition, we observe that the performance of the IMA-GNN architecture can increase linearly with an increase in the number of resistive CAM and MVM crossbars in decentralized setting for various datasets and saturate once the entire node feature data could be fitted onto the crossbars. However, it comes at the cost of higher power consumption for each node.

% finding motivates the added benefit of a hybrid, semi-decentrlaized setting outlined in \cite{nazzal2023traffic}, where one bridges the computation-communication tradeoff 

% finding is consistent with the arguments in \cite{nazzal2023traffic} about this trade off.

% the centralized setting has obvious benefits over the decentralized GNN setting in terms of communication latency, while in terms of computation latency, the decentralized setting shows a much higher performance compared to the centralized setting, especially in large graph networks. This result is consistent with the argument in \cite{nazzal2023traffic}

% This result conforms with the
% This reveals the need for a hybrid semi-decentralized GNN accelerator design in the future.

\section{Conclusion}
While the respective benefits of centralized and decentralized GNNs are known in software implementation, there is a lack of hardware implementation analysis to show the communication and computation loads in each setting. This work undertakes this task by modeling and analyzing practical case studies on GNN-based taxi demand and supply prediction and adopting large-scale graph datasets. Our cross-layer simulation results demonstrate our proposed platform called IMA-GNN in the centralized GNN setting can obtain $\sim$790$\times$ communication speed-up compared to the decentralized GNN setting. However, the decentralized GNN setting performs computation $\sim$1400$\times$ faster while reducing the power consumption per device. 
This study is conducted based on certain assumptions as discussed. Nevertheless, the results extrapolate that the decentralized GNN setting achieves gains in reducing the computation latency. However, this comes at the expense of increasing communication overhead and latency. This latency is more strongly pronounced with larger graphs. This finding confirms the necessity and the potential of balancing this communication-computation trade-off through a semi-decentralized setting \cite{nazzal2023traffic}. In that setting, multiple edge devices are employed to decentralize the operation on a graph level while each edge device region works in a centralized fashion. Our future work will consider the hardware acceleration of the semi-decentralized GNN setting.

\bibliographystyle{IEEEtran}
\bibliography{sample-base}

\end{document}